# Highly Sensitive and Robust Ultrahigh-Q Nanocavity on a Multiheterostructure Photonic Crystal

**Ashfaqul Anwar Siraji**
Wayne State University, Department of Electrical and Computer Engineering, Detroit, Michigan, USA, 48202

**Abstract**. An ultrahigh-Q hybrid cavity using a graded multiheterostructure and space modulation is designed in this work. Two high Q resonant modes exist in the cavity and follow separate confinement mechanisms. We have demonstrated an analytical method for designing the desired multiheterostructure. Space modulation was used to create a robust and sensitive nanocavity. The confinement mechanism and resonant characteristics of the resonant modes are studied using the finite difference time domain method. We have shown that the modes of the cavity are highly robust with respect to fabrication error. We have also demonstrated that the nanocavity can act as a highly sensitive refractive index sensor.

## 1   Introduction

Planar photonic crystal (PPC) based optical nanocavities based on photonic crystal offer extremely high spectral purity, which makes them useful for sensing applications, lasers etc[1-5]. Among the different planar photonic crystal nanocavities, the double heterostructure (DH) nanocavities have garnered special attention for the past few years. This is because DH nanocavities offer several advantages over defect based photonic crystal cavities. They have high Q, can be easily incorporated into planar photonic structures and are highly robust[7]. DH nanocavities have been demonstrated by creating a localized perturbation the PhC lattice surrounding a line defect[6-9]. Recently we have demonstrated a tunable nanocavity based on rectangular lattice that can sustain both TE and TM modes[9].

Several ultrahigh-Q nanocavities using the idea gradual perturbation of the PhC lattice have been demonstrated[10,11]. A PC cavity in bulk GaN using a multiheterostructure (MH) that uses four successive photonic crystal lattices has been demonstrated[10]. A ultrahigh-Q PhC cavity consisting of many successive photonic crystal lattices[11] have also been demonstrated. In such MH nanocavities, the confinement along the line defect waveguide is carefully controlled to obtain gradual spatial confinement.

In this work, we propose a hybrid nanocavity which utilizes multiheteostructure in combination with space modulation[13]. We use two dimensional (2D) finite difference time domain (FDTD) method with perfectly matched layer (PML) as the boundary conditions to calculate the properties of the resonant modes of the hybrid nanocavity. In the 2D geometry, we study the TE modes. The effect of space modulation on the MH nanocavity is studied. Furthermore, we study the robustness of the resonant modes with respect to fabrication error on the resonant modes of the hybrid cavity by calculating the resonant wavelengths after introducing random position disorder in the structure. Finally, the possible application of the designed cavity in refractive index sensing is discussed.

## 2  Design and Verification

We design the MH-PC by extending the principle demonstrated in[11]. We consider a line defect waveguide corresponding to a row of missing hole along the X axis in a hexagonal lattice PC with lattice constant *a* and airhole radius $r = 0.35 \times a$. The background material is assumed to be Si *(n = 3.4)*. The dispersion curve of the TE mode of the waveguide is determined by eigenvalue calculation using 2D-FDTD assuming finite width and infinite length. In Fig. 1, the dispersion curve of the guided mode in terms of normalized frequency versus the wavevector along X axis, *k* is shown. For simplicity, the subscript (x) is dropped. The slope of this dispersion curve near the mode-edge is almost zero, indicating bandgap guiding as opposed to index guiding. The dispersion curve is fitted by a Taylor series expansion of the term *(k-0.5)* where *k* is the wavevector.

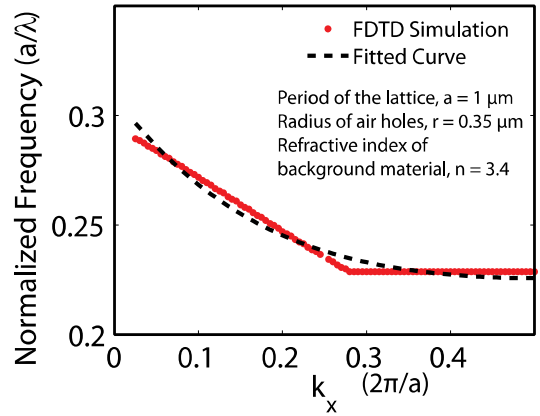

**Fig. 1** The calculated dispersion curve of a hexagonal lattice PC waveguide and its best fitted function. The lattice constant of the PC, *a* = 1 µm, airhole radius r = 0.35 × *a*, the background material is silicon (n = 3.4) and the waveguide corresponds to a row of filled up airhole.

The fitted function, displayed in Fig. 1, can be expressed as:
$$f = 0.2257 + 0.1541(k-0.5)^2 + 0.7048(k-0.5)^4 + H.O.T., \tag{1}$$
where *H.O.T* stands for higher order terms. Since the guiding mechanism of the waveguide is bandgap guiding, the dispersion is not quadratic, unlike that presented in Ref. 11. We consider terms upto the fourth power, since the fit is sufficiently close without terms of higher order. By substituting *k = (0.5+iq)* in (1), we can obtain the complex dispersion relation in the bandgap region, viz.,
$$f = 0.2257 - 0.1541q^2 + 0.7048q^4, \tag{2}$$
where *q* is the imaginary part of the wavevector. In order to obtain a Gaussian resonant mode profile along the X axis, the condition on *q* is
$$q = Bx, \tag{3}$$
where B is an arbitrary constant. From (2) and (3), we can obtain
$$f_{cut} - f = 0.1541B^2x^2 - 0.7148B^2x^2, \tag{4}$$

Here, $f_{cut}$ is the cutoff frequency of the waveguide and $f$ is the resonant frequency of the cavity. Since the resonant frequency of the cavity should be constant, the $x$ dependent term in (4) is $f_{cut}(x)$. Hence, $f_{cut}$ should be different in different regions of the PC. The cutoff frequency and lattice constant of a PC is inversely proportional to each other. Hence,

$$\frac{f_{cut}(x)}{f_{cut}(0)} = \frac{a(0)}{a(x)}, \tag{5}$$

where $f_{cut}(0) = \frac{0.2257c}{a}$ is the cutoff frequency at $x = 0$ (central region of the MH cavity) and $a(0)$ is the lattice constant in the same region. The resonant modes of a heterostructure cavity form near the cutoff frequency of the dispersion curve[11]. Therefore, we assume that the resonant frequency of the cavity will be close to $f = f_{cut}(0)$. Then, $a(x)$ can be written as

$$a(x) = \frac{a(0)f_{cut}(0)}{f_{cut}(0) + 0.1541B^2 x^2 - 0.7048B^4 x^4}. \tag{6}$$

Assuming small change in the lattice constant $(\delta n)$ and two periods per PC, the distance $(x_n)$ of the of the $n^{th}$ PC from the center of the MH cavity is $x_n = (2n+0.5)a(0)\$$. Using this relationship with (6), we obtain

$$a_n = \frac{a(0)f_{cut}(0)}{f_{cut}(0) + 0.1541B^2 a(0)^2 (2n+0.5)^2 - 0.7048B^4 a(0)^4 (2n+0.5)^4}, \tag{7}$$

which produces the lattice constant of the $n^{th}$ PC away from the center. In (7), the constant $B$ is arbitrary. We select a value for $B$ such that the variation in lattice constant remains within 3% of $a(0)$, so that our assumption of small $\delta n$ remains valid.

We have designed the multiheterostructure (MH) nanocavity using (7). The MH consists of a 15 successive photonic crystals (PC) with gradually changing lattice constants along the X axis. The central PC, denoted by $PC_0$ has the largest lattice constant. The two PCs on either side of $PC_0$, denoted by $PC_1$, have smaller lattice constants than that of $PC_0$. The two PCs on left and right of two $PC_1$s, respectively, denoted by $PC_2$, have smaller lattice constants than those of the $PC_1$s. This way, the successive PCs away from $PC_0$ have gradually smaller lattice constants, dictated by (7). The designed MH nanocavity is shown in Fig. 2(a). The $PC_0$ and $PC_1$s are highlighted in the figure. The bandgap of the PC also changes gradually as the lattice constant changes along the X axis. As shown in Fig. 2(b), the bandgap is smallest in the $PC_0$ region where the lattice constant is largest. As the lattice constant becomes gradually smaller on either side of $PC_0$, the bandgap becomes larger.

The normalized impulse response of the MH nanocavity for TE mode is shown in Fig. 3(a). It can be seen that the resonant peak $(\lambda_r)$ is at $c/a = 0.24$, which is very close to the $f_{cut}(0) = 0.2257 (c/a)$ as obtained using the analytical procedure. In Fig. 3(b), we show the decay of energy in the nanocavity with respect to time when it is excited at its resonant frequency. The quality factor (Q) of this MH nanocavity has been calculated using the method outlined in Ref. 9 to be $8.9 \times 10^4$. It is compared with Q from previous literature in Table 1. It can be seen that the Q of this nanocavity compares favourably with previously reported values.



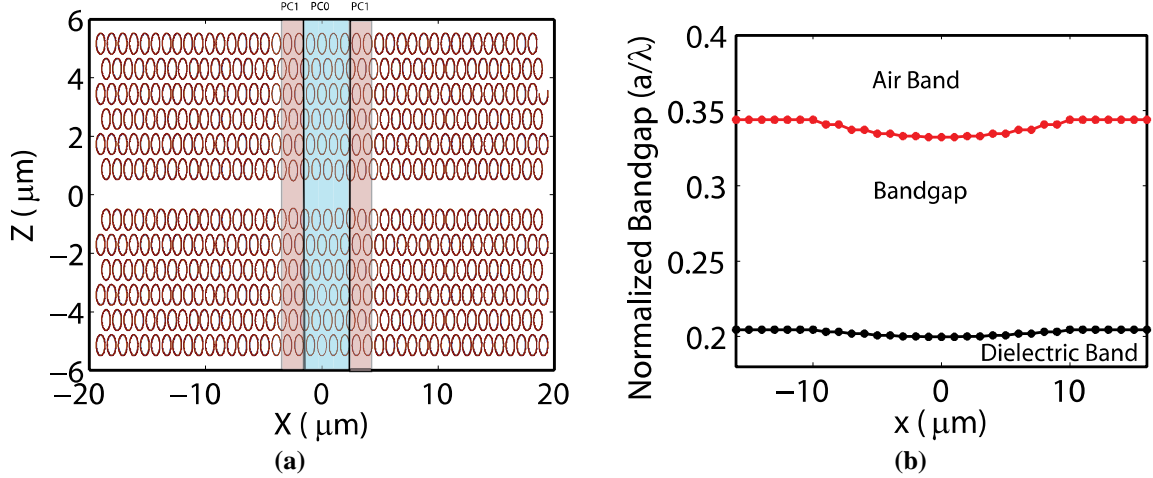

**Fig. 2** (a) The structure of the designed MH nanocavity with $PC_0$ and $PC_1$ highlighted, (b) the gradual change in the bandgap of the designed MH nanocavity.

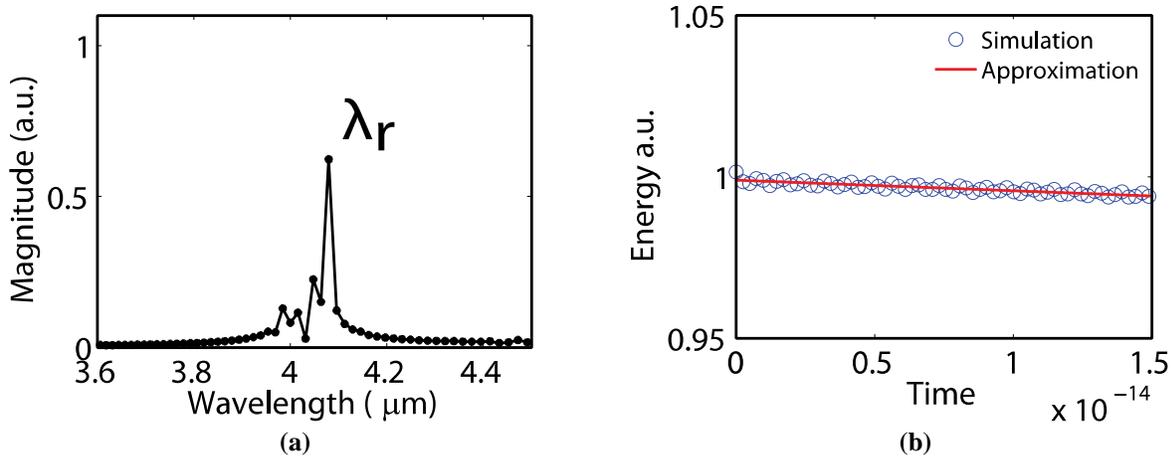

**Fig. 3** (a) The impulse response of the MH nanocavity. The resonant wavelength is at $a/\lambda = 0.24$. (b) The decay of energy with respect to time in the nanocavity when excited by the resonant frequency.

**Table 1** Comparison of quality factor of the MH nanocavity with previously reported values.

| References | Material | $Q_{in-plane}$ |
|---|---|---|
| Tanaka et al[11] | Si | $10^9$ |
| Mock et al[14] | Si | $3.37 \times 10^5$ |
| Takahashi et al[12] | Si | $2.5 \times 10^6$ |
| This work | SI | $8.9 \times 10^4$ |

The resonant TE mode profile of the MH nanocavity and its spatial Fourier transform (SFT) are shown in Figs. 4(a) and 4(b), respectively. The mode profiles show gradual confinement in space and sharp confinement in the momentum domain. For comparison, the mode profiles of a L3 defect nanocavity[13] and a DH nanocavity[9] are shown in Fig. 5. Since the three nanocavities are of different



sizes, we calculated the ratio of the mode area to the device area for the three cavities. We have found that the ratio is 0.94% for an MH nanocavity, 0.37% for a DH nanocavity and 0.34% for the L3 nanocavity. As expected, the MH nanocavity has a more gradual spatial confinement compared to that of the DH nanocavity, whereas the confinement is sharpest for an L3 cavity. Such gradual confinement in the MH nanocavity results in a Q factor which is several orders of magnitude higher.

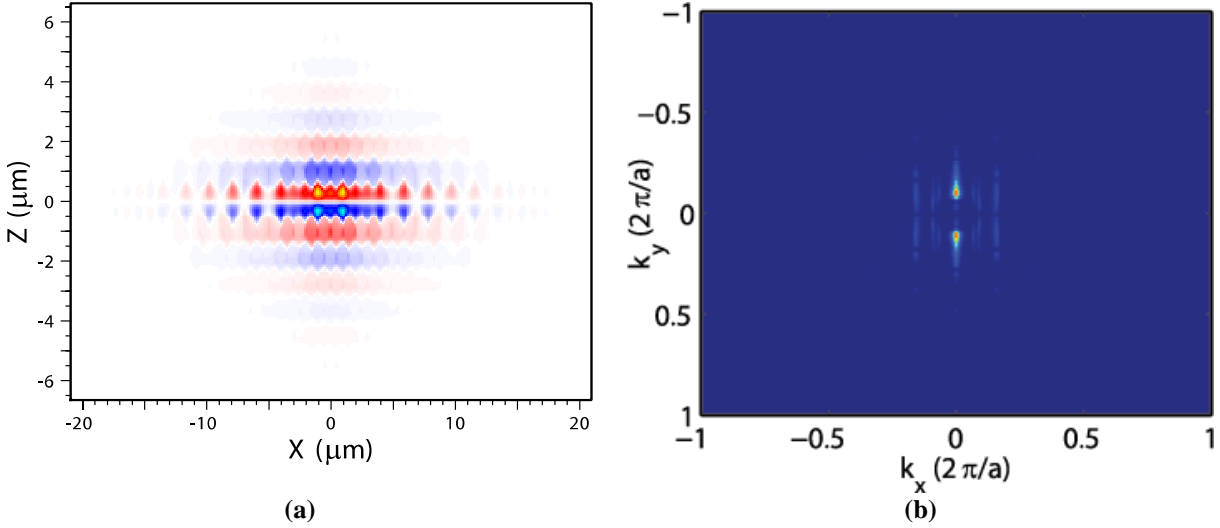

**Fig. 4** (a) The resonant mode profile of the MH cavity, (b) The SFT of the mode

We have employed (7), that assumes a Gaussian envelope for the resonant mode, to design the MH nanocavity. We now verify that the resonant mode of the designed MH nanocavity has a Gaussian envelope. The magnetic field profiles of the resonant TE mode along X and Z axes are shown in Fig.6. Using the formalism used in Ref. 9, we can calculate the least error fit of the envelope of magnetic field as:

$$H = e^{-(ax^m + bx^{m+1})}, \qquad (8)$$

where $H$ is the envelope of the magnetic field, $m$ is the performance parameter and $a,b$ are fitting parameters. We first calculate $m$ so that the fitting error is minimized. In the inset of Fig.6, we show the RMS fitting error against $m$. It can be seen that along the X axis, the least error fit of the envelope of the magnetic field is given by $H_x = A\ exp(-ax^2)$, with $A$ being a constant. This is the desired Gaussian envelope. However, along the Z axis, the least error fit of the envelope requires $H_z = A exp(-az^4)$, indicating a sharper spatial confinement.



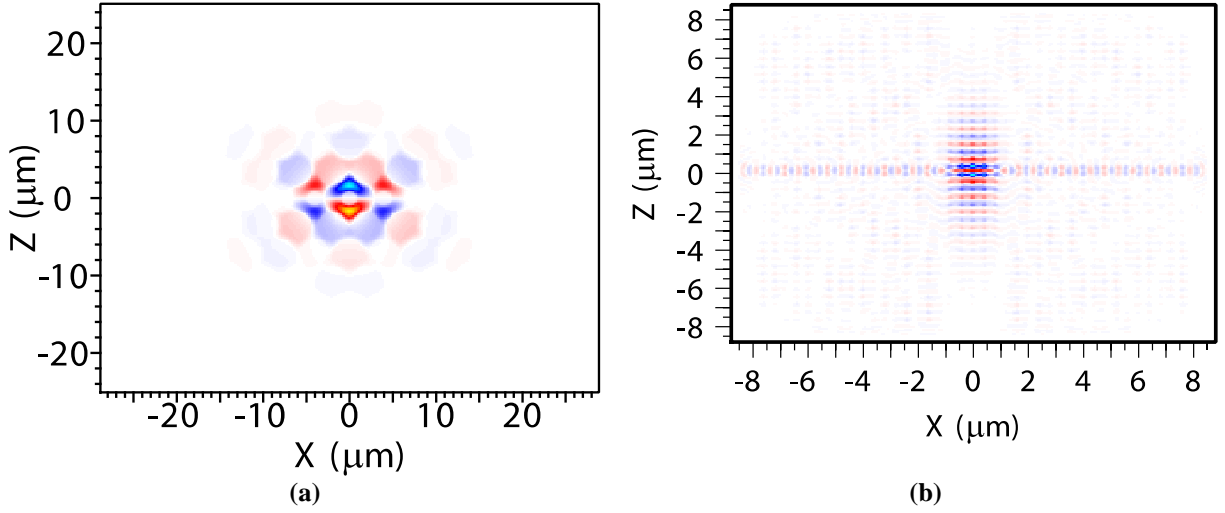

**Fig. 5** (a) The resonant mode of a L3 type nanocavity where Q = $4 \times 10^3$. (b) The resonant mode of a DH nanocavity where Q = $2.8 \times 10^3$.

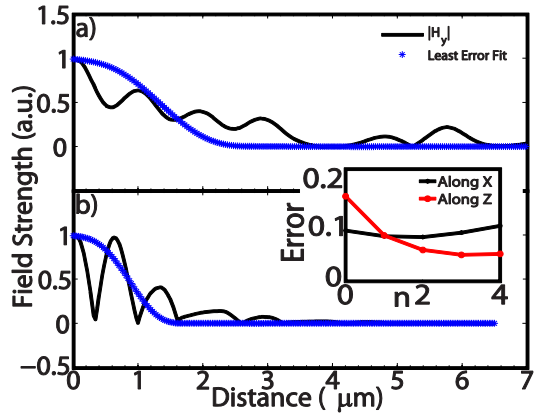

**Fig. 6** The resonant magnetic field profile of the MH cavity and the least error fit of its envelope along the (a) X axis and (b) Z axis. The inset shows the RMS fitting error plotted against the performance parameter *m*.

## 3  Impact of Space Modulation

The MH nanocavity designed in the previous section has resonant magnetic field profile with Gaussian envelope along X axis as well as a very high Q. However, the confinement is still very sharp along the Z axis. As can be seen from the resonant mode profile, the line defect can be thought of as a series of L1 cavity. Displacing the airholes immediately around such cavities can produce softer transition between cavity and mirror, producing higher Q[13,15]. Hence we apply space modulation along Z axis to the airholes immediately around the MH nanocavity. We calculated the frequency response of the MH nanocavity after applying space modulation. The normalized impulse response is shown in Fig. 7. It can be seen that there are two resonant peaks at λ = 3.82 μm and 3.83 μm, respectively. The profiles of the magnetic field at the two resonant wavelengths are shown in Fig.8. Using the method used in the previous section, we calculated the least error fit for the envelope of the magnetic fields along Z axis for both modes. In case of the first mode, it has the form *H = Aexp(-az)* while in case of the second mode, it has the form *H = Aexp(-az$^2$)*.



Thus, by using space modulation, light can be confined much more gradually along the Z axis. The Q of the mode at 3.82 μm was calculated to be $1.2 \times 10^5$, which is 34.83% higher than the Q of the unmodulated cavity.

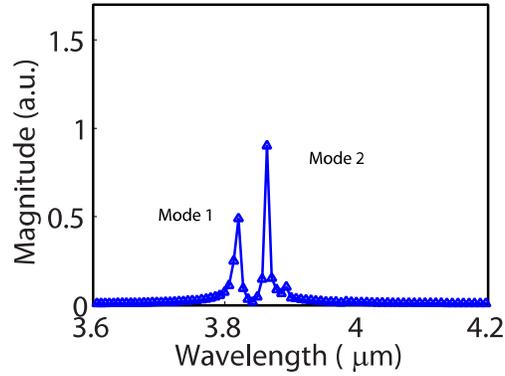

**Fig. 7** The normalized impulse responses of the MH cavity after applying space modulation.

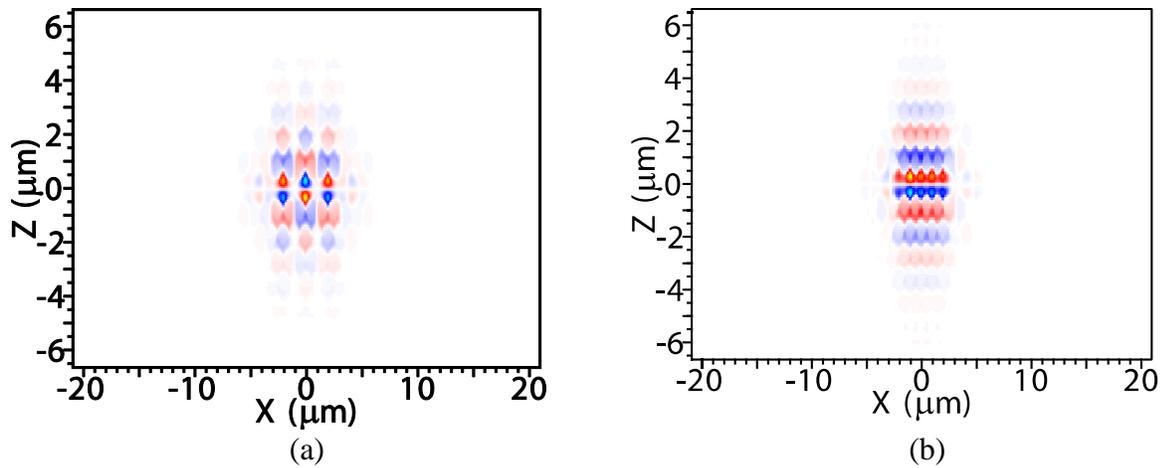

**Fig. 8** The magnetic field profile at the two resonant frequencies, (a) mode 1 at $a/\lambda = 0.2617$, and (b) mode 2 at $a/\lambda = 0.2588$.

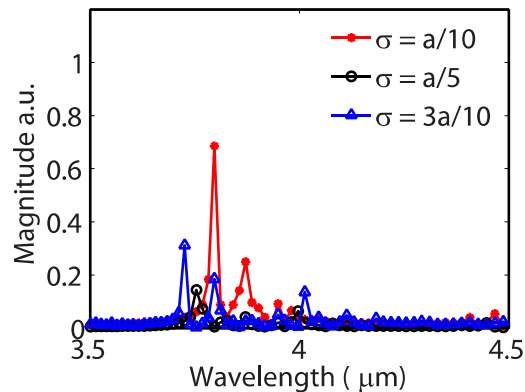

**Fig. 9** The normalized impulse response of the cavity with added disorder. Results are shown for disorders of increasing standard deviation.



## 4 Disorder Stability

Most of the times, due to fabrication error some uncertainty on the design parameters of a PC cavity is introduced. For example, all the air holes may not be of same radius or all the airholes may not be in the exact position specified by the design. Defect cavities are specially sensitive to this type of errors. In Ref. 16, three types of defect cavity and three types of DH cavity were fabricated and it was found that the resonant wavelengths change significantly with increasing fabrication disorder. To investigate the disorder stability of the designed hybrid nanocavity, we calculated the impulse response of the nanocavity after introducing disorder in the positions of the airholes. We performed separate calculations for added position disorder with standard deviation (σ) of $\sigma = a/10$, $\sigma = a/5$ and $\sigma = 3a/10$ respectively. The results are shown in the Fig. 9. From the figure, it is evident that despite increasing position disorder, the hybrid nanocavity continues to demonstrate two separate resonant modes whose wavelengths remain relatively unchanged. The results are also summarized in Table 2, where the standard deviation of the added position disorder (σ) is shown along with the percentage deviation in the two resonant wavelengths. The wavelength of the first resonant mode ($\lambda_1$) is slightly less robust than that of the second resonant mode ($\lambda_2$). As σ varies within 30% of the lattice constant (*a*), the wavelength of the first resonant mode ($\lambda_1$) changes by as much as 2.51%, whereas the wavelength of the second resonant mode ($\lambda_2$) changes by only 1.734%. This can be traced back to the confinement mechanism of these two modes.

Table 2 Percent change in the resonant wavelengths of the cavity.

| Standard Deviation | $\Delta\lambda_1$ | $\Delta\lambda_2$ |
|---|---|---|
| 0.1*a* | 0.62% | 0.181% |
| 0.2*a* | 1.78% | 3.44% |
| 0.3*a* | 2.51% | 1.734% |

The first mode of the hybrid cavity behaves more like a resonant mode of a donor defect cavity, which are especially vulnerable to position disorder[16]. But the second mode acts more like a MH cavity, which shows remarkable robustness against position disorder[6].

## 5 Refractive Index Sensing

To investigate the refractive index sensing capabilities of the designed cavity, the frequency response of the cavity for different background index were calculated. As shown in Fig. 10(a), the resonant peak of the cavity is gradually red shifted for higher refractive index. In Fig. 10(b), the position of the resonant peak is plotted against increasing refractive index. The data obtained from simulation is approximated by a linear equation as:

$$\lambda_{res} = C + S \times n_{BG}, \tag{9}$$

where *C = 2.5154* µm is a constant and *S = 884.25 nm/RIU* is the average sensitivity. In Table 3, the sensitivity and quality factor of this work is compared with those in recent literature. It can be seen that the MH cavity produces significantly higher sensitivity and Q.

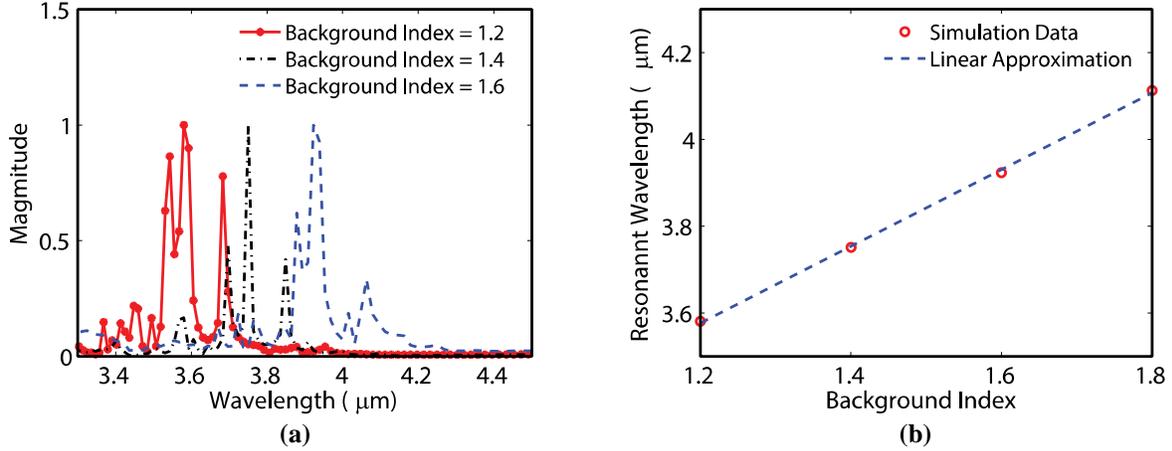

**Fig. 10** (a) The frequency response of the designed cavity with different background index. It can be observed that the resonance shifts to higher wavelength when the background index is higher. (b) Linear fit of the calculated resonance peak for increasing background index.

**Table 3** Comparison between the refractive index sensitivity found in this work and that in similar works in literature.

| Reference | Material | Sensitivity (nm/RIU) | Q |
|---|---|---|---|
| Feng et al[17] | Si | 190 | $10^6$ |
| Falco et al[18] | SOI | 585 | $5 \times 10^4$ |
| This work | Si | 884 | $1.2 \times 10^5$ |

## 6 Conclusion

In this work, we have demonstrated an analytical design flow for designing MH cavity with Gaussian field profile even when the waveguide dispersion is non-quadratic. The space modulation is introduced along the direction perpendicular to the waveguide in the MH, which resulted in a higher Q and lower mode area. The resonant mode with smaller wavelength acts like a donor defect cavity while the resonant mode with larger wavelength acts like a MH cavity. These resonant modes have different resonant field profiles but similar high values of Q. The wavelengths of the two resonant modes of this hybrid cavity becomes closer to each other at higher modulation depth. But the Q of the modes monotonously decrease at higher modulation depth, though the value never fall below the Q of unmodulated MH cavity. We have also investigated the disorder stability of the cavity by simulating the cavities with added position disorder. We have found that both the modes of the hybrid cavity are highly insensitive to disorder in position. Since this cavity can sustain two separate high Q modes with stability against fabrication disorder, this type of cavities can be used in lasers, switching and in planar photonic circuits.